\documentclass{iopart}
\usepackage{euscript,iopams,amssymb,amsfonts,graphicx,bm}
\usepackage{pgfplots,setstack}

\usepackage{float}
\usepackage{epsfig}
\usepackage{esint}
\usepackage{color}

\usepackage{bm,braket}

\eqnobysec

\newcommand{\q}{\widetilde{q}}

\newcommand{\p}{\widetilde{p}}

\newcommand{\R}{{\mathbb R}}

\renewcommand{\e}{\mathrm{e}}

\newcommand{\sgn}{\mathrm{sgn}}

\newcommand{\x}{\mathbf{x}}
\newcommand{\y}{\mathbf{y}}

\newcommand{\X}{\mathbf{X}}
\renewcommand{\P}{\mathbb{P}}

\begin{document}
\title[Accumulation time of diffusion processes with resetting]{Accumulation time of stochastic processes with resetting}
\author{Paul C. Bressloff}
\address{Department of Mathematics, University of Utah, Salt Lake City, UT, USA} \ead{bressloff@math.utah.edu}

\begin{abstract}
One of the characteristic features of a stochastic process under resetting is that the probability density converges to a nonequilibrium stationary state (NESS). In addition, the approach to the stationary state exhibits a dynamical phase transition, which can be interpreted as a traveling front separating spatial regions for which the probability density has relaxed to the NESS from those where it has not. Since the trajectories contributing to the transient region are rare events, one can establish the existence of the phase transition by carrying out an asymptotic expansion of the exact solution. In this paper we develop an alternative, direct method for characterizing the approach to the NESS of a stochastic process with resetting that is based on the calculation of the so-called accumulation time. The latter is the analog of the mean first passage time of a search process, in which the survival probability density is replaced by an accumulation  fraction density. In the case of one-dimensional Brownian motion with Poissonian resetting, we derive the asymptotic formula $|x-x_0|\approx \sqrt{4rD}T(x)$ for $|x-x_0|\gg\sqrt{D/r}$, where $T(x)$ is the accumulation time at $x$, $r$ is the constant resetting rate, $D$ is the diffusivity and $x_0$ is the reset point. This is identical in form to the traveling front condition for the dynamical phase transition. We also derive an analogous result for diffusion in higher spatial dimensions and for non-Poissonian resetting. We then consider the effects of delays such as refractory periods and finite return times. In both cases we establish that the asymptotic behavior of $T(x)$ is independent of the delays. Finally, we extend the analysis to a run-and-tumble particle with resetting. We thus establish the accumulation time of a stochastic process with resetting as a useful quantity for characterizing the approach to an NESS (if it exists) that is relatively straightforward to calculate.  
\end{abstract}

\maketitle

\section{Introduction}

A topic of increasing interest  within the statistical physics community is the theory of stochastic processes under
resetting. The simplest example of such a process is a Brownian particle whose position is reset randomly in time at a constant rate $r$ (Poissonian resetting) to some fixed point $\x_r$, which is often identified with its initial position $\x_0$ \cite{Evans11a,Evans11b,Evans14}. One of the characteristic features of diffusion under stochastic resetting is that the probability density converges to a nonequilibrium stationary state (NESS) that maintains nonzero probability currents. In addition, one finds that the approach to the stationary state exhibits a dynamical phase transition, which can be interpreted as a traveling front separating spatial regions for which the probability density has relaxed to the NESS from those where it has not. Since the trajectories contributing to the transient region are rare events, one can establish the existence of the phase transition by carrying out an asymptotic expansion of the exact solution. The existence of a nontrivial NESS has also been established for a wider range of stochastic processes with resetting, including non-diffusive processes such as Levy flights \cite{Kus14} and run-and-tumble processes \cite{Evans18,Bressloff20}, switching diffusions \cite{Bressloff20a}, non-Poissonian resetting \cite{Eule16,Pal16,Nagar16}, and diffusion in a potential landscape \cite{Pal15}.  Another extension has been the inclusion of delays due to finite return times \cite{Mendez19,Pal19,Pal19a,Bodrova20,Pal20} and refractory periods \cite{Reuveni14,Evans19a,Mendez19a}.

In this paper we develop an alternative method for characterizing the approach to the NESS of a diffusion process with resetting that is based on the calculation of the so-called accumulation time. The latter is the analog of the mean first passage time of a search process, in which the survival probability density is replaced by an accumulation  fraction density. Accumulation times are commonly used to estimate the time to form a protein concentration gradient during morphogenesis in order to check that gradient formation is consistent with developmental time scales \cite{Berez10,Berez11,Gordon11,Bressloff19}; the non-trivial stationary concentration gradient is maintained by a local source of protein synthesis at the boundary of the domain combined with absorption. We begin by calculating the accumulation time for one-dimensional (1D) Brownian motion with instantaneous Poissonian resetting (section 2). In particular, we derive the asymptotic formula $|x-x_0|\approx \sqrt{4rD}T(x)$ for $|x-x_0|\gg\sqrt{D/r}$, where $T(x)$ is the accumulation time at $x$, $r$ is the constant resetting rate, $D$ is the diffusivity and $x_0$ is the reset point. This is formally identical to the traveling front condition for the dynamical phase transition obtained in \cite{Majumdar15}. We also derive an analogous result for diffusion in higher spatial dimensions and with non-Poissonian resetting. We then consider the effects of delays such as refractory periods and finite return times (section 3). In both cases we establish that the asymptotic behavior of the accumulation time is independent of the delays. One of the advantages of quantifying the approach to the NESS in terms of the accumulation time is that it is relatively straightforward to apply to other types of stochastic process. We illustrate this in the case of a run-and-tumble particle with resetting (section 4).

\setcounter{equation}{0}
\section{Brownian particle in $\R^d$ with resetting}
Consider a single Brownian particle diffusing in $\R^d$ and resetting at a rate $r$ to its initial position $\x_0$ \cite{Evans11a,Evans11b,Evans14}. In an infinitesimal time interval $dt$,
\begin{equation}
\X(t+dt)=\left \{ \begin{array}{ll} \x_0 & \mbox{ with probability } rdt\\
\X(t)+\sqrt{2D}d{\bf W}(t) & \mbox{ with probability } (1-rdt)\end{array}\right . ,
\end{equation}
where ${\bf W}(t)$ is a $d$-dimensional Wiener process with independent components. Let $p(\x,t)$ be the probability density for the particle to be at position $\x$ at time $t$. The forward differential Chapman-Kolmogorov (CK) equation takes the form
\begin{equation}
\label{reset}
\frac{\partial p(\x,t)}{\partial t}=D\nabla^2 p(\x,t)-rp(\x,t)+r\delta(\x-\x_0),
\end{equation}
with the initial condition $p(\x,0)=\delta(\x-\x_0)$. It is usually more convenient to work with a renewal equation \cite{Evans14,Evans20}. In the absence of resetting ($r=0$), equation (\ref{reset}) reduces to pure diffusion so that $p=p_0(\x,t)$ where $p_0$ is the fundamental solution or propagator:
\begin{equation}
p_0(\x,t)=\frac{1}{(4\pi Dt)^{d/2}}\e^{-|\x-\x_0|^2/4Dt}.
\end{equation}
When resetting is included, the probability density $p(\x,t)$ has two distinct types of contribution: paths where no resetting events have occurred up to time $t$, and paths where the last resetting event occurred at time $\tau_l=t-\tau$ for some $\tau \in (0,t)$. In the case of Poissonian resetting with rate $r$, the probability density of no resetting events up to time $t$ is $\e^{-rt}$. Similarly, the probability density that the last resetting event occurred at time $t-\tau$ (with no subsequent resetting events) is $r\e^{-r\tau}$. Using the fact that in the latter case $X(t-\tau)=x_0$ and one has pure diffusion over the time interval $(t-\tau,t)$, the full time-dependent solution to the CK equation (\ref{reset}) satisfies the so-called last renewal equation
\begin{equation}
\label{renewal}
p(\x,t)=\e^{-rt}p_0(\x,t)+r\int_0^tp_0(\x,\tau)\e^{-r\tau}d\tau.
\end{equation}
 A complementary version is the so-called first renewal equation
\begin{equation}
\label{renewal2}
p(\x,t)=\e^{-rt}p_0(\x,t)+r\int_0^tp(\x,t-\tau)\e^{-r\tau}d\tau.
\end{equation}
The second term now integrates over all trajectories for which a first reset to $x_0$ occurred between time $\tau$ and $\tau+d\tau$ followed by possibly multiple further resetting events in the remaining time $t-\tau$.
One of the useful features of these renewal equations is that they hold for more general stochastic processes by 
taking $p_0$ to be the appropriate propagator. 

Laplace transforming the renewal equation (\ref{renewal}) gives
\begin{equation}
\label{renewalLT}
\widetilde{p}(\x,s)=\widetilde{p}_0(\x,s+r)+\frac{r}{s}\widetilde{p}_0(\x,r+s).
\end{equation}
It follows that the steady-state density is
\begin{equation}
\label{pss}
p^*(\x)=\lim_{s\rightarrow 0} s\widetilde{p}(\x,s)=r\widetilde{p}_0(\x,r).
\end{equation}
Note that $p^*(\x)$ represents a non-equilibrium stationary state (NESS) because there exist non-zero probability fluxes. 
For $d=1$, we have
\begin{equation}
\label{prop1D}
\widetilde{p}_0(x,s)=\frac{1}{2\sqrt{sD}}\e^{-\sqrt{s/D}|x-x_0|}.
\end{equation}
and
\begin{equation}
\label{pss1D}
p^*(x)=\frac{1}{2}\sqrt{\frac{r}{D}}\e^{-\sqrt{r/D}|x-x_0|}.
\end{equation}
For arbitrary $d$, the Laplace transformed propagator is given by a modified Bessel function \cite{Evans14}:
\begin{equation}
\label{prop2D}
\widetilde{p}_0(\x,s)= \frac{1}{2\pi D} \left (2\pi \sqrt{\frac{D}{s}} |\x-\x_0| \right )^{\nu}K_{\nu}(\sqrt{s/D}|\x-\x_0|),
\end{equation}
with $\nu=1-d/2$. Hence,
\begin{equation}
p^*(\x)= \frac{r}{2\pi D} \left (2\pi\sqrt{\frac{D}{r}} |\x-\x_0| \right )^{\nu}K_{\nu}(\sqrt{r/D}|\x-\x_0|).
\end{equation}
Using the identity
\begin{equation}
K_{-1/2}(y)=K_{1/2}(y)=\sqrt{\frac{\pi}{2y}}\e^{-y},
\end{equation}
we recover the $d=1$ result and for $d=3$ we have
\begin{equation}
p^*(\x)= \frac{r}{4\pi |\x-\x_0|}\e^{-\sqrt{r/D}|\x-\x_0|}.
\end{equation}
One finds that in a neighborhood of $\x_0$, $p^*(\x)\sim \ln|\x-\x_0|$ for $d=2$ and $p^*(\x)\sim 1/|\x-\x_0$ for $d=3$.

\subsection{Accumulation time}

In order to construct the accumulation time for $p$ to reach steady-state, consider the function
\begin{equation}
\label{Z}
Z(\x,t)=1-\frac{p(\x,t)}{p^*(\x)},
\end{equation}
which represents the fractional deviation of the concentration from the steady-state. Assuming that there is no overshooting, $1-Z(\x,t)$ is the fraction of the steady-state concentration that has accumulated at $\x$ by time $t$. It follows that $-\partial_t Z(\x,t)dt$ is the fraction accumulated in the interval $[t,t+dt]$. The accumulation time is then defined by analogy to mean first passage times, \cite{Berez10,Berez11,Gordon11}
\begin{equation}
T(\x)=\int_0^{\infty} t\left (-\frac{\partial Z(\x,t)}{\partial t}\right )dt=\int_0^{\infty} Z(\x,t)dt.
\end{equation}
Note that a finite accumulation time implies that the steady-state is a stable solution. 
It is usually more useful to calculate an accumulation time in Laplace space. Using the identity 
\[p^*(\x)=\lim_{t\rightarrow \infty} p(\x,t)=\lim_{s\rightarrow 0}s\widetilde{p}(\x,s),\]
and setting $\widetilde{F}(\x,s)=s\widetilde{p}(\x,s)$, the Laplace transform of equation (\ref{Z}) gives
\[s\widetilde{Z}(\x,s)=1-\frac{\widetilde{F}(\x,s)}{\widetilde{F}(\x,0)},\]
and, hence
\begin{eqnarray}
\fl T(\x)&=&\lim_{s\rightarrow 0} \widetilde{Z}(\x,s) = \lim_{s\rightarrow 0}\frac{1}{s}\left [1-\frac{\widetilde{F}(\x,s)}{\widetilde{F}(\x,0)}\right ] =-\frac{1}{\widetilde{F}(\x,0)}
\left .\frac{d}{ds}\widetilde{F}(\x,s)\right |_{s=0}.
\label{acc}
\end{eqnarray}
Using the Laplace transformed renewal equation (\ref{renewalLT}),
\begin{eqnarray}
f(\x)\equiv  \left .\frac{d}{ds}\widetilde{F}(\x,s)\right |_{s=0}&=\left .\frac{d}{ds}\left (s\widetilde{p}_0(\x,s+r)+r\widetilde{p}_0(\x,r+s)\right )\right |_{s=0}\nonumber \\
&=\widetilde{p}_0(\x,r)+r\partial_r\widetilde{p}_0(\x,r),
\label{f}
\end{eqnarray}
which implies that
\begin{equation}
\label{accu2}
T(\x)=-\frac{1}{r}-\frac{\partial_r\widetilde{p}_0(\x,r)}{\widetilde{p}_0(\x,r)}.
\end{equation}

Note that one derive an alternative expression for $f(\x)$ by considering the differential equation for $\widetilde{F}(\x,s)$. Laplace transforming the CK equation (\ref{reset}) and multiplying both sides by $s$ gives
\begin{equation}
\label{resetLT}
D\nabla^2 \widetilde{F}(\x,s)-(r+s)\widetilde{F}(\x,s)=-(r+s)\delta(\x-\x_0).
\end{equation}
Taking the limit $s\rightarrow 0$ recovers the time-independent equation for $p^*(\x)$.
On the other hand, differentiating both sides with respect to $s$ and then taking $s\rightarrow 0$ with $f(\x)= \partial_s \widetilde{F}(\x,s)|_{s=0}$ we have
\begin{equation}
\label{resetLT2}
D\nabla^2  {f}(\x)-r\ {f}(\x)=-\delta(\x-\x_0)+p^*(\x).
\end{equation}
This equation can be solved in terms of the propagator $\p_0(\x,r)$ according to
\begin{equation}
\label{falt}
f(\x)=\p_0(\x,r) -r\int_{\R^d} \p_0(\x-\y,r)\p_0(\y,r)d\y.
\end{equation}
after using the result $p^*(\x)=r\p_0(\x,r)$.
It can be shown that (\ref{falt}) is equivalent to (\ref{f}) by noting that $\partial_r\widetilde{p}_0(\x,r)$ satisfies the equation
\begin{equation}
D\nabla^2 \partial_r\widetilde{p}_0(\x,r)-r\partial_r\widetilde{p}_0(\x,r)=\widetilde{p}_0(\x,r).
\end{equation}

In the 1D case, the propagator is given by equation (\ref{prop1D}) so that
\begin{equation}
\label{T1D}
T(x)=\frac{1}{2r}\left [\sqrt{\frac{r}{D}}|x-x_0|  -1\right ].
\end{equation}
For locations close to the restart point $x_0$, $|x-x_0|<\sqrt{D/r}$, we see that $T(x)<0$ which is a consequence of the probability density overshooting in the presence of resetting. This effect becomes negligible when $|x-x_0|\gg\sqrt{D/r}$ such that
\begin{equation}
\label{front1D}
\sqrt{4rD}T(x)\approx |x-x_0|\gg \sqrt{\frac{D}{r}}.
\end{equation}
Equation (\ref{front1D}) is identical in form to the result derived in Ref. \cite{Majumdar15} by carrying out an asymptotic expansion of the exact solution for large $t$ and using steepest descents. This suggests that $T(x)$ represents the position of a traveling front that effectively separates the spatial regions for which the probability density has relaxed to the NESS from those where it has not. The accumulation time provides a more direct method for obtaining the front condition and also yields a sufficient condition for its validity. Moreover, it is easily generalizable to other stochastic processes with resetting, as we show in this paper for a range of examples. It should be noted, however, that one limitation of using the accumulation time is that it cannot establish the sharpness of the boundary between transient and NESS regions. This would require higher-order statistics of the fractional accumulation density, for example.

An analogous result to equation (\ref{front1D}) can be derived in higher spatial dimensions by using the propagator (\ref{prop2D}). In particular,
\begin{eqnarray}
\fl \partial_r\widetilde{p}_0(\x,r)&=-\frac{1}{2\pi D} \left (2\pi \sqrt{D/r} |\x-\x_0| \right )^{\nu}\bigg \{K_{\nu}(\sqrt{r/D}|\x-\x_0|)\frac{\nu}{r}  \\
\fl &  \qquad +\frac{1}{2}\sqrt{\frac{1}{rD}} |\x-\x_0|  K_{\nu-1}( \sqrt{r/D} |\x-\x_0| ) \nonumber 
\bigg \}.
\end{eqnarray}
We have used the Bessel identity
\begin{equation}
K_{\nu}'(z)=-K_{\nu-1}(z)-\frac{\nu}{z}K_{\nu}(z).
\end{equation}
The corresponding accumulation time is thus
\begin{equation}
T(\x)=\frac{\nu-1}{r}+\frac{1}{2}\sqrt{\frac{1}{rD}} |\x-\x_0|  \frac{K_{\nu-1}( \sqrt{r/D} |\x-\x_0| ) }{K_{\nu}(\sqrt{r/D}|\x-\x_0|)}.
\end{equation}
The front condition for large $|\x-\x_0|$ is of the form
\begin{equation}
\label{front2D}
\sqrt{4rD}T(\x)\approx |\x-\x_0|  \frac{K_{\nu-1}( \sqrt{r/D} |\x-\x_0| ) }{K_{\nu}(\sqrt{r/D}|\x-\x_0|)} \approx |\x-\x_0|,
\end{equation}
since $|\x-\x_0|\gg \sqrt{D/r}$ and $K_{\nu}(z)\sim \sqrt{\pi/2z}\e^{-z}$ for large $z$.

\subsection{Non-Poissonian resetting}

So far we have assumed that resetting occurs at a constant rate $r$ (Poissonian resetting). A more general resetting protocol is to take the sequence of resetting times to be generated by a probability density $\psi(\tau)$ \cite{Eule16,Pal16,Nagar16}. It follows that $\Psi(\tau)=1-\int_0^{\tau}\psi(s)ds$ is the probability that no resetting has occurred up to time $\tau$. 
 For Poissonian resetting we have $\psi(\tau)=r\e^{-r\tau}$ and $\Psi(\tau)=\e^{-r\tau}$.
The first renewal equation (\ref{renewal2}) easily generalizes as
\begin{equation}
p(x,t)=\Psi(t)p_0(x,t)+\int_0^t\psi(\tau)p(x,t-\tau)d\tau.
\end{equation}
Taking the Laplace transform of this equation yields
\begin{equation}
\p(x,s)=\int_0^{\infty}\e^{-st}\Psi(t) p_0(x,t)dt+ \widetilde{\psi}(s) \p(x,s).
\end{equation}
Rearranging and using the identity
\begin{equation}
\widetilde{\Psi}(s)=\frac{1-\widetilde{\psi}(s)}{s},
\end{equation}
shows that
\begin{equation}
\p(x,s)=\frac{1}{s\widetilde{\Psi}(s)}\int_0^{\infty}\e^{-st}\Psi(t) p_0(x,t)dt.
\end{equation}
Multiplying by $s$ and taking the limit $s\rightarrow 0$ determines the stationary density (assuming it exists)
\begin{equation}
p^*(x)=\frac{\int_0^{\infty}\Psi(t)p_0(x,t)dt}{\int_0^{\infty}\Psi(t)dt}.
\end{equation}
A sufficient condition for existence of $p^*(x)$ is that
\begin{equation}
\int_0^{\infty} \Psi(t)dt<\infty,
\end{equation}
which requires that $\psi(t)$ decays faster than $1/t^2$. In order to determine the accumulation time when $p^*(x)$ exists, we consider the derivative
\begin{eqnarray}
\fl \partial_s[s\p(x,s)]=- \frac{\widetilde{\Psi}'(s)}{\widetilde{\Psi}(s)}\int_0^{\infty}\e^{-st}\Psi(t)p_0(x,t)dt-\frac{1}{\widetilde{\Psi}(s)}\int_0^{\infty}t\e^{-st}\Psi(t) p_0(x,t)dt.
\end{eqnarray}
It follows from equation (\ref{acc}) that the accumulation time is
\begin{equation}
\label{Tnon}
T(x)=-\widetilde{\Psi}'(0)+\frac{\int_0^{\infty}t\Psi(t)p_0(x,t)dt}{\int_0^{\infty}\Psi(t)p_0(x,t)dt},
\end{equation}
with
\begin{equation}
p_0(x,t)=\frac{1}{\sqrt{4\pi Dt}}\e^{-|x-x_0|^2/4Dt}.
\end{equation}

Unfortunately, the calculation of the accumulation time can no longer be carried out using Laplace transforms. However, it is still possible to derive a front condition using steepest descents. For the sake of illustration, we follow \cite{Pal16} and consider 1D diffusion with a time-dependent resetting rate for which
\begin{equation}
\psi(t)=r(t)\e^{-R(t)},\quad \Psi(t)=\e^{-R(t)},\quad R(t)=\int_0^tr(\tau)d\tau,
\end{equation}
and $\int_0^{\infty}\e^{-R(t)}dt <\infty$. In particular, suppose that $r(t)=b_0t^{\theta}$ with $\theta >-1$ and thus \cite{Pal16}
\begin{equation}
R(t)=\frac{b_0t^{1+\theta}}{1+\theta}.
\end{equation} 
For large $|x-x_0|$, $T(x)$ in equation (\ref{Tnon}) is dominated by the integral terms, which can be evaluated using steepest descents. That is,
\begin{eqnarray}
T(x)\approx \frac{\int_0^{\infty}t^{1/2}\e^{-S(x,t)}dt}{\int_0^{\infty}t^{-1/2}\e^{-S(x,t)}dt},
\end{eqnarray}
where 
\begin{equation}
S(x,t)=\frac{b_0t^{1+\theta}}{1+\theta}+\frac{|x-x_0|^2}{4Dt}.
\end{equation}
The integrals are dominated by times in a neighborhood of $t_*$ with $\partial_tS(x,t_*)=0$:
\begin{eqnarray}
b_0t_*^{\theta}-\frac{|x-x_0|^2}{4Dt_*^2}=0
\end{eqnarray}
or
\begin{equation}
t^*=\left (\frac{|x-x_0|^2}{4Db_0}\right )^{1/(2+\theta)}.
\end{equation}
It follows that to leading order,
\begin{eqnarray}
T(x)\approx \frac{t_*^{1/2} \e^{-S(x,t_*)}}{t_*^{-1/2} \e^{-S(x,t_*)}}=t_*.
\end{eqnarray}
Combining the previous two equations yields the asymptotic behavior
\begin{equation}
(4b_0D)^{1/(2+\theta)}T(x)\approx |x-x_0|^{1/(1+\theta/2)}.
\end{equation}
This has the same form as the traveling front condition derived in \cite{Pal16} by performing an asymptotic expansion of the last renewal equation for $p(x,t)$. In the case of Poissonian resetting ($\theta=0$) with $b_0=r$ we recover equation (\ref{front1D}).

\section{Stochastic resetting with delays}

\subsection{Effect of a refractory period}
Now suppose that whenever the particle returns to $\x_0$, it is subject to a refractory period before reentering the diffusion state \cite{Reuveni14,Evans19a,Mendez19a}. The refractory period is itself a random variable with a corresponding waiting time density $W$, which is taken to have a finite mean $\langle \tau\rangle $ and second moment $\tau_2$. Following the particular formulation of \cite{Evans19a}, the inclusion of a refractory period into the first renewal equation (\ref{renewal2}) yields
\begin{eqnarray}
p(\x,t )&=\e^{-rt}p_0(\x,t)+r\int_0^tdt'\e^{-rt'}\int_0^{t-t'}d\tau \ W(\tau)p(\x,t-t'-\tau)\nonumber\\
 &\quad +r\int_0^tdt'\e^{-rt'}\int_{t-t'}^{\infty} d\tau \ W(\tau)\delta(\x-\x_0).
\end{eqnarray}
The first term on the right-hand side is the contribution from trajectories without resetting; the second term integrates over the first resetting time $t'$, which is followed by a refractory period $\tau$ so that the particle takes a time $t-t'-\tau$ to reach $\x$; the final term includes trajectories that first reset at $t'$ and are still in the refractory state at time $t$. Laplace transforming the modified renewal equation using the convolution theorem and then rearranging gives \cite{Evans19a}
\begin{eqnarray}
\fl \p(\x,s)=\frac{1}{r+s-r\widetilde{W}(s)}\left [(r+s)\p_0(\x,r+s)+\frac{r}{s}[1-\widetilde{W}(s)]\delta(x)\right ].
\end{eqnarray}
Multiplying both sides by $s$ and taking the limit $s\rightarrow 0$ with
\begin{equation}
\label{Wtil}
\widetilde{W}(s)\approx 1-s\langle \tau\rangle+s^2\tau_2/2,
\end{equation}
we obtain the NESS \cite{Evans19a}
\begin{equation}
p^*(\x)=\frac{r}{1+r\langle \tau\rangle}[\p_0(\x,r)+\delta(\x-\x_0)\langle \tau\rangle ].
\end{equation}
(If the mean $\langle \tau \rangle$ diverges then $p^*(\x)\rightarrow \delta(\x-\x_0)$.)

The relaxation of the component multiplying the delta function peak was analyzed in \cite{Evans19a}. Here we focus on the approach to the stationary state for $\x\neq \x_0$. In order to determine the accumulation time we need to evaluate the derivative
$\partial_s s\p(\x,s)$:
\begin{eqnarray}
\fl \partial_s [s\p(\x,s)]&=\partial_s \left [\frac{s(r+s)}{r+s-r\widetilde{W}(s)}\p_0(\x,r+s)\right ]\nonumber \\
\fl &=\left [\frac{r+2s}{r+s-r\widetilde{W}(s)}-\frac{s(r+s)[1-r\widetilde{W}'(s)]}{[r+s-r\widetilde{W}(s)]^2}\right ]\p_0(\x,r+s)  \nonumber \\
\fl &\qquad +\frac{s(r+s)}{r+s-r\widetilde{W}(s)}\partial_s \p_0(\x,r+s).
\end{eqnarray}
Substituting for $\widetilde{W}(s)$ using (\ref{Wtil}) and taking the limit $s\rightarrow 0$ leads to the result
\begin{eqnarray}
\fl \partial_s [s\p(\x,s)]&=\left [\frac{1}{1+r\langle \tau\rangle}+\frac{r^2\tau_2}{2[1+r\langle \tau\rangle]^2}\right ]\p_0(\x,r)  +\frac{r}{1+r\langle \tau\rangle}\partial_s \p_0(\x,r).
\end{eqnarray}
Equation (\ref{acc}) then implies that for $\x\neq \x_0$
\begin{equation}
T(\x)=-\frac{1}{r}- \frac{r\tau_2}{2[1+r\langle \tau\rangle ]}-\frac{\partial_r \p_0(\x,r)}{ \p_0(\x,r)}.
\end{equation}
Comparison with equation (\ref{accu2}) for the accumulation time without a refractory period shows that the only effect of refractoriness on the accumulation time is to modify the overshoot at locations $\x$ close to $\x_0$. For example, in 1D we now have
\begin{equation}
\label{T1Dref}
T(x)=\frac{1}{2r}\left [\sqrt{\frac{r}{D}}|x-x_0|  -1-\frac{r^2 \tau_2}{1+r\langle \tau \rangle}\right ].
\end{equation}
Sufficiently far from the initial point $x_0$, the accumulation time exhibits the same asymptotic behavior as found for no refractory period, and equation (\ref{front1D}) becomes
\begin{equation}
\label{front1Dref}
\sqrt{4rD}T(x)\approx |x-x_0|\gg \sqrt{\frac{D}{r}}\left [1+\frac{r^2\tau_2}{1+r\langle \tau \rangle}\right ].
\end{equation}
One possible interpretation of the invariance of the asymptotic accumulation time with respect to refractoriness is that, although a larger mean refractory time $\langle \tau \rangle$ implies that the particle spends a larger fraction of its time at $\x_0$ rather than diffusing, there is less probability to be accumulated at $\x\neq \x_0$.

\subsection{Effect of a finite return time} Another possible source of delay in the resetting process is a finite return time \cite{Mendez19,Pal19,Pal19a,Pal20,Bodrova20}. Rather than instantaneously returning to $\x_0$ following reset, suppose that the particle switches to a ballistic state in which it returns to $\x_0$ at a constant speed $v_0$. (More general return dynamics are considered in \cite{Pal19a,Bodrova20}.) Following \cite{Pal19}, we focus on the 1D case and take $x_0=0$ so that the velocity of return is $v_0$ for $x>0$ and $-v_0$ for $x<0$. The probability density $p(x,t)$ can be decomposed as
\begin{equation}
p(x,t)=p_D(x,t)+p_R(x,t),
\end{equation}
where $p_D(x,t)$ and $p_R(x,t)$ represent the contributions from the diffusive motion phase and the ballistic return phase, respectively. We also have the marginal probabilities
\begin{equation}
P_D(t)=\int_{-\infty}^{\infty}p_D(x,t)dx,\quad P_R(t)=\int_{-\infty}^{\infty}p_R(x,t)dx.
\end{equation}
The two components satisfy the evolution equations \cite{Pal19,Bodrova20}
\numparts
\begin{equation}
\label{pD}
\frac{\partial p_D(x,t)}{\partial t} =D\frac{\partial^2p_D(x,t)}{\partial x^2} -rp_D(x,t)+2\delta(x)v_0p_R(0,t)
\end{equation}
and
\begin{equation}
\label{pR}
\frac{\partial p_R(x,t)}{\partial t} =\sgn(x)v_0\frac{\partial p_R(x,t)}{\partial x}+rp_D(x,t).
\end{equation}
\endnumparts
Equation (\ref{pD}) is a modified version of (\ref{reset}) in which the total probability flux associated with instantaneously reentering the diffusion phase at $x=0$ has equal contributions from the left-moving and right-moving ballistic fluxes arriving at the origin. Equation (\ref{pR}) is the Louiville equation for deterministic drift combined with the fact that switching from the diffusion to the ballistic phase occurs at a rate $r$.

Rather than constructing a renewal equation in order to determine the accumulation times, we directly Laplace transform the evolution equations:
\numparts
\begin{equation}
\label{pDLT}
D\frac{\partial^2\p_D(x,s)}{\partial x^2}-(r+s)\p_D(x,s)= -[2v_0\p_R(0,s)+1]\delta(x)
\end{equation}
and
\begin{equation}
\label{pRLT}
\sgn(x)v_0\frac{\partial \p_R(x,s)}{\partial x}-s\p_R(x,s)+r\p_D(x,s)=0.
\end{equation}
\endnumparts
These equations can be solved explicitly to give \cite{Pal19}
\numparts
\begin{equation}
\label{pDsol}
\p_D(x,s)=\frac{1}{2s}\frac{s+v_0\sqrt{(r+s)/D}}{v_0+\sqrt{D(r+s)}}\e^{-\sqrt{(r+s)/D}|x|}
\end{equation}
and
\begin{equation}
\label{pRsol}
\p_R(x,s)=\frac{1}{2s}\frac{r}{v_0+\sqrt{D(r+s)}}\e^{-\sqrt{(r+s)/D}|x|}.
\end{equation}
\endnumparts
As highlighted by Pal {\em et al.} \cite{Pal19}, the total probability density is independent of the return speed, and is thus identical to the result for 1D diffusion with instantaneous resetting
\begin{equation}
\fl \p(x,s)=\p_R(x,s)+\p_D(x,s) =\p_R(x,s)=\frac{1}{2s}\sqrt{\frac{r+s}{D}}\e^{-\sqrt{(r+s)/D}|x|}.
\end{equation}
It follows that the total stationary density and the approach to stationarity are identical to the instantaneous case. (This is analogous to the invariance of the phase transition with respect to refractory periods.) However, the individual components have distinct stationary states and accumulation times.

Multiplying equations (\ref{pDsol}) and (\ref{pRsol}) by $s$ and taking the limit $s\rightarrow 0$ yields
\numparts
\begin{equation}
\label{pDss}
p_D^*(x)=\frac{1}{2}\sqrt{\frac{r}{D}}\frac{v_0}{v_0+\sqrt{rD}}\e^{-\sqrt{r/D}|x|}=\frac{1}{2}\sqrt{\frac{r}{D}}P_D^*\e^{-\sqrt{r/D}|x|}.
\end{equation}
and
\begin{equation}
p_R^*(x)=\frac{1}{2}\sqrt{\frac{r}{D}}\frac{\sqrt{rD}}{v_0+\sqrt{rD}}\e^{-\sqrt{r/D}|x|}=\frac{1}{2}\sqrt{\frac{r}{D}}P_R^*\e^{-\sqrt{r/D}|x|},
\end{equation}
\endnumparts
where $P_D^*$ and $P_R^*$ are the stationary probabilities of being in the diffusive and ballistic phases, respectively. Note that $P_D^*+P_R^*=1$ such that $P_D^*\rightarrow 0$ when $v_0\rightarrow 0$ and $P_R^*\rightarrow 0$ when $v_0\rightarrow \infty$.
In addition,
\numparts
\begin{eqnarray}
\fl \partial_s [s\p_D(x,s)]&= \bigg \{\frac{1+v_0/2\sqrt{D(r+s)}}{s+v_0\sqrt{(r+s)/D}}-\frac{\sqrt{D/(r+s)}/2}{v_0+\sqrt{(r+s)D}}-\frac{|x|}{2\sqrt{D(r+s)}}\bigg \}s\p_D(x,s)\nonumber \\
\fl 
\end{eqnarray}
and
\begin{eqnarray}
\fl \partial_s [s\p_R(x,s)]&= -\bigg \{\frac{\sqrt{D/(r+s)}/2}{v_0+\sqrt{(r+s)D}}+\frac{|x|}{2\sqrt{rD}}\bigg \}s\p_R(x,s).\end{eqnarray}
\endnumparts
Introducing the component accumulation times
\begin{eqnarray}
\fl T_{D,R}(x)&=\int_0^{\infty}\left [1-\frac{p_{D,R}(x,t)}{p_{D,R}^*(x)}\right ]dt=-\frac{1}{\widetilde{F}_{D,R}(x,0)}
\left .\frac{d}{ds}\widetilde{F}_{D,R}(x,s)\right |_{s=0}
\end{eqnarray}
with $\widetilde{F}_{D,R}(x,s)=s\p_{D,R}(x,s)$, we obtain the results (assuming $0<v_0<\infty$)
\numparts
\begin{equation}
\label{TD}
T_D(x)=\frac{1}{2r}\left [-\frac{2\sqrt{rD}+v_0}{v_0}+\frac{\sqrt{rD}}{v_0+\sqrt{rD}}+\sqrt{\frac{r}{D}}|x|\right ]
\end{equation}
and
\begin{equation}
T_R(x)=\frac{1}{2r}\left [\frac{\sqrt{rD}}{v_0+\sqrt{rD}}+\sqrt{\frac{r}{D}}|x|\right ].
\end{equation}
\endnumparts
We see that for sufficiently large $|x|$ both components have the same asymptotic behavior as found in the case of instantaneous resetting. Moreover, we recover equation (\ref{T1D}) by taking the limit $v_0\rightarrow \infty$ in equation (\ref{TD}). On the other hand, for finite $v_0$, the regime where $T_D(x)$ exhibits front-like behavior requires $|x|\gg D/v_0$, which diverges as $v_0\rightarrow 0$.

It turns out that the steady-state result (\ref{pDss}) for an appropriately defined $P_D^*$ also holds for more general forms of return dynamics such as space-dependent return speeds $v(x)$ \cite{Pal19a}. That is, the spatial variation of the probability density for the stochastic motion in the diffusive phase is independent of the velocity profile $v(x)$.  In order to explore this issue from the perspective of the accumulation time, consider the analog of equations (\ref{pDLT}) and (\ref{pRLT2}) for $v=v(x)$:
\numparts
\begin{equation}
\label{pDLT2}
D\frac{\partial^2\p_D(x,s)}{\partial x^2}-(r+s)\p_D(x,s)= -[2v(0)\p_R(0,s)+1]\delta(x)
\end{equation}
and
\begin{equation}
\label{pRLT2}
\sgn(x)\frac{\partial v(x)\p_R(x,s)}{\partial x}-s\p_R(x,s)+r\p_D(x,s)=0.
\end{equation}
\endnumparts
Integrating the second equation with respect to $x$ shows that
\begin{equation}
\p_R(0,s)= \frac{r\widetilde{P}_D(s)-s\widetilde{P}_R(s)}{2v(0)}=\frac{(r+s)\widetilde{P}_D(s)-1}{2v(0)},
\end{equation}
where $\widetilde{P}_D(s)=\int \p_D(x,s)dx$. 
Substituting into (\ref{pDLT2}) then gives
\begin{equation}
\label{pDLT3}
D\frac{\partial^2\p_D(x,s)}{\partial x^2}-(r+s)\p_D(x,s)= -(r+s)\widetilde{P}_D(s)\delta(x).
\end{equation}
Using a similar argument to \cite{Pal16}, this has a solution of the form 
\begin{equation}
\p_D(x,s)=s\widetilde{P}_D(s)\p_{\infty}(x,s),
\end{equation}
where
$\p_{\infty}(x,s)$ is the Laplace transform of the probability density for instantaneous resetting:
\begin{equation}
D\frac{\partial^2\p_{\infty}(x,s)}{\partial x^2}-(r+s)\p_{\infty}(x,s)= -\frac{r+s}{s}\delta(x).
\end{equation}
It immediately follows that
$p_D^*(x)=P^*_Dp_{\infty}^*(x)$, where $p_{\infty}^*(x)$ is the total stationary density for instantaneous resetting. In addition,
\begin{eqnarray}
f_D(x)\equiv \left .\frac{d}{ds}\widetilde{F}_D(x,s)\right |_{s=0}=P_D^* f_{\infty}(x)+p^*_{\infty}(x)\int_{-\infty}^{\infty} f_D(y)dy
\end{eqnarray}
so that
\begin{eqnarray}
T_D(x)=T_{\infty}(x)-\frac{1}{P_D^*}\int_{-\infty}^{\infty} f_D(y)dy.
\end{eqnarray}
Hence, the accumulation time in the diffusion phase is equal to the accumulation time under instantaneous resetting but shifted by a constant. This is what we found explicitly for constant speed $v_0$, see equation (\ref{TD}).

\section{Run-and-tumble particle}

Consider a run-and-tumble particle (RTP) that randomly switches between two constant velocity states labeled by $n=\pm$ with $v_+=v$ and $v_-=-v$ for some $v>0$. Furthermore, suppose that the particle reverses direction according to a Poisson process with rate $\alpha$. The position $X(t)$ of the particle at time $t$ then evolves according to the piecewise deterministic equation
\begin{equation}
\label{PDMP}
\frac{dX}{dt}=v\sigma(t),
 \end{equation}
where $\sigma(t)=\pm 1$ is a dichotomous noise process that switches sign at the rate $\alpha$. Let $q_{\sigma}(x,t)$ be the probability density of the RTP at position $x\in \R$ at time $t>0$ and moving to the right ($\sigma=1)$ and to the left ($\sigma=-1$), respectively. The associated CK equation is then
\numparts
\begin{eqnarray}
\label{DLa}
\frac{\partial q_{1}}{\partial t}&=-v \frac{\partial q_{1}}{\partial x}-\alpha q_{1}+\alpha q_{-1},\\
\frac{\partial q_{-1}}{\partial t}&=v \frac{\partial q_{-1}}{\partial x}-\alpha q_{-1}+\alpha q_{1}.
\label{DLb}
\end{eqnarray}
\endnumparts
This is supplemented by the initial conditions $x(0)=x_0$ and $\sigma(0)=\sigma_0=\pm 1$ with probability $\rho_{\pm 1}$ such that $\rho_1+\rho_{-1}=1$.

Now suppose that the position $X(t)$ is reset to its initial location $x_0$ at random times distributed according to an exponential distribution with rate $r\geq 0$ \cite{Evans18}. The evolution of the system over the infinitesimal time $dt$ is then
\numparts
\label{X}
\begin{equation}
X(t+dt)=\left \{ \begin{array}{ccc}X(t)+v\sigma(t)dt & \mbox{with probability} & 1-rdt\\
x_0 & \mbox{with probability} & rdt,
\end{array} \right .
\end{equation}
and
\begin{equation}
\sigma(t+dt)=\left \{ \begin{array}{c}\sigma(t) \, \mbox{with probability} \, 1-rdt-\alpha dt\\
-\sigma(t) \, \mbox{with probability} \, \alpha dt   \\
\sigma_0=\pm 1 \, \mbox{with probability} \, r\rho_{\pm1}dt.
\end{array} \right .
\end{equation}
\endnumparts
The resulting probability density with resetting, which we denote by $p_{n}$, evolves according to the modified CK equation \cite{Evans18} 
\numparts
\label{CKH2}
\begin{eqnarray}
\frac{\partial p_{1}}{\partial t}&=-v \frac{\partial p_{1}}{\partial x}-(\alpha +r) p_{1}+\alpha  p_{-1}
+r\delta(x-x_0)\rho_1,\\
\frac{\partial p_{-1}}{\partial t}&=v \frac{\partial p_{-1}}{\partial x}-(\alpha +r) p_{-1}+\alpha  p_{1}
+r\delta(x-x_0)\rho_{-1}.
\end{eqnarray}
\endnumparts
In Ref. \cite{Evans18}, the NESS was determined in the symmetric case $\rho_{\pm}=1/2$ and $x_0=0$ by noting the the total density with resetting, $p= p_{1}+p_{-1}$, is related to the corresponding total density without resetting, $q=q_0+q_1$, according to a last renewal equation identical in form to (\ref{renewal}):
\begin{eqnarray}
p(x,t)=\e^{-rt}q(x,t)+r\int_0^t\e^{-r\tau}q(x,\tau)d\tau.
\end{eqnarray}
The analysis of section 2.1 thus carries over to a RTP with resetting. In particular, working in Laplace space we find that the NESS is
\begin{equation}
p^*(x)=r\widetilde{q}(x,r),
\end{equation}
and the associated accumulation time is
\begin{equation}
\label{accRTP}
T(x)=-\frac{1}{r}-\frac{\partial_r\widetilde{q}(x,r)}{\widetilde{q}(x,r)}.
\end{equation}
The function $\q(x,r)$ can be calculated by Laplace transforming equations (\ref{DLa}) and (\ref{DLb}), and one finds that \cite{Evans18}
\begin{equation}
\q(x,r)=\frac{\lambda}{2r}\e^{-\lambda |x|},\quad \lambda=\lambda(r)\equiv \sqrt{\frac{r(r+2\alpha)}{v^2}}.
\end{equation}
Substituting into equation (\ref{accRTP}) gives
\begin{eqnarray}
\fl T(x)=\lambda'(r)\left [-\frac{1}{\lambda(r)}+|x|\right ]=-\frac{r+\alpha}{r(r+2\alpha)}+\frac{r+\alpha}{\sqrt{r(r+2\alpha)}}\frac{|x-x_0|}{v_0}.
\end{eqnarray}
Hence, for large $|x-x_0|$ we have the asymptotic behavior
\begin{equation}
v_0\frac{\sqrt{r(r+2\alpha)}}{r+\alpha}T(x)\approx |x-x_0|.
\end{equation}
Note that in the fast switching limit, $\alpha \rightarrow \infty$, 
\begin{equation}
\sqrt{\frac{4v_0^2r}{\alpha}}T(x)\sim |x-x_0|. 
\end{equation}
Comparison with equation (\ref{front1D}) shows that $v_0^2/2\alpha$ acts as an effective diffusivity, consistent with the well-known diffusion limit of an RTP. Similar results can be obtained when $\rho_1\neq \rho_{-1}$ \cite{Bressloff20}.

\section{Conclusion}

In this paper we analyzed the approach to the NESS of Brownian motion in $\R^d$ with instantaneous Poissonian resetting by calculating the accumulation time $T(\x)$. We showed that for $|\x-\x_0|\gg \sqrt{D/r}$, the accumulation time varies as $\sqrt{4Dr}T(\x)\sim |\x-\x_0|$, which has the form of a traveling front that is consistent with a dynamical phase transition. We generalized this result by considering non-Poissonian resetting, delays due to refractory periods or finite return times, and a run-and-tumble particle. 

It would be interesting to explore in more detail the connection (if any) between the asymptotic behavior of the accumulation time and the dynamical phase transition of the full probability density based on large deviation theory \cite{Majumdar15}. Although the accumulation time is easy to define and calculate, it doesn't itself provide evidence of a second-order phase transition.

\end{document}